\documentclass{article}
\pdfoutput=1

\usepackage{arxiv}
\usepackage[utf8]{inputenc} 
\usepackage[T1]{fontenc} 
\usepackage{hyperref} 
\hypersetup{
	colorlinks = true,
	linkcolor = black,
	anchorcolor = black,
	citecolor = black,
	filecolor = black,
	urlcolor = blue
}
\usepackage{booktabs} 
\usepackage{amsfonts} 
\usepackage{nicefrac} 
\usepackage{microtype} 
\usepackage{amsmath,bm}
\usepackage{graphicx}
\usepackage[flushleft]{threeparttable}
\usepackage{multirow}
\usepackage{natbib}
\usepackage{makecell}
\usepackage{caption}
\usepackage{subcaption}   
\usepackage{float}

\newcommand{\figurehere}[1]{\begin{center}%
		=========================\\%
		Insert Figure #1 about here\\%
		=========================\\%
\end{center}}
\newcommand{\tablehere}[1]{\begin{center}%
		=========================\\%
		Insert Table #1 about here\\%
		=========================\\%
\end{center}}

\usepackage{array}
\newcommand{\PreserveBackslash}[1]{\let\temp=\\#1\let\\=\temp}
\newcolumntype{C}[1]{>{\PreserveBackslash\centering}p{#1}}
\newcolumntype{R}[1]{>{\PreserveBackslash\raggedleft}p{#1}}
\newcolumntype{L}[1]{>{\PreserveBackslash\raggedright}p{#1}}

\title{Extending Latent Basis Growth Model to Explore Joint Development in the Framework of Individual Measurement Occasions}

\author{
  Jin Liu \thanks{CONTACT Jin Liu Email: Veronica.Liu0206@gmail.com}\\
}

\begin{document}
\maketitle
\begin{abstract}
Longitudinal processes often pose nonlinear change patterns. Latent basis growth models (LBGMs) provide a versatile solution without requiring specific functional forms. Building on the LBGM specification for unequally-spaced waves and individual occasions proposed by \citet{Liu2022LCSM}, we extend LBGMs to multivariate longitudinal outcomes. This provides a unified approach to nonlinear, interconnected trajectories. Simulation studies demonstrate that the proposed model can provide unbiased and accurate estimates with target coverage probabilities for the parameters of interest. Real-world analyses of reading and mathematics scores demonstrates its effectiveness in analyzing joint developmental processes that vary in temporal patterns. Computational code is included.
\end{abstract}

\keywords{Latent Basis Growth Model \and Parallel Nonlinear Longitudinal Processes \and Individual Measurement Occasions \and Simulation Studies}

\section{Introduction}\label{sec:Intro}
In longitudinal studies, researchers often gather measurements on multiple outcomes to understand how each evolves over time. While most studies have focused on univariate outcomes, real-world processes in domains like development \citep{Shin2013RM, Liu2021PBLSGM, Peralta2020PBLSGM}, behavioral sciences \citep{Duncan1994substances, Duncan1996substances}, and biomedicine \citep{Dumenci2019knee}, often interrelate. This is reflected in some recent works that consider multiple longitudinal outcomes. Researchers in these studies aim not just to gauge within-individual changes and between-individual differences of each process but also to explore how changes in multiple outcomes are interconnected. Various research contexts exemplify this. For instance, developmental studies often collect achievement scores in multiple subjects \citep{Shin2013RM, Liu2021PBLSGM, Peralta2020PBLSGM, Liu2022PBLSGMM}, enabling a comprehensive analysis of how progress in one area correlates with another. Similarly, clinical trials may collect multiple endpoints \citep{Dumenci2019knee} to provide a holistic evaluation of treatment effects. 

In another potential scenario, researchers may focus on evaluating how well different data sources agree over the course of the study. For instance, child and parent reports could both provide data on a child's health-related quality of life in observational studies \citep{Rajmil2013Multi}. In clinical trials, it is common for a single endpoint to be measured using different machines. There is also interest in studying repeated outcomes from distinct individuals who are nested within pairs or small groups \citep{McNulty2016multi, Lyons2017multi}. The objective of our study is to develop a model within the Structural Equation Modeling (SEM) framework. This model aims to describe the joint nonlinear trajectories of either two longitudinal outcomes or a univariate repeated outcome from multiple sources. It also aims to estimate the degree of association between these variables.

When it comes to modeling these longitudinal outcomes, capturing accurate trajectory shapes is crucial, especially when the data show nonlinear change patterns over time \citep{Cudeck2007Nonlinear}. A range of parametric models, such as polynomial, exponential, and logistic growth curves, provide alternatives to capture these nonlinear characteristics. In addition, prior work has successfully utilized piecewise models, which employ semi-parametric functions like linear-linear and linear-quadratic piecewise, to depict complex trajectories with varying rates of change \citep{Harring2006nonlinear, Flora2008knot, Dumenci2019knee, Kohli2011PLGC, Kohli2013PLGC1, Kohli2013PLGC2, Kohli2015PLGC1, Kohli2015PLGC2, Liu2021PBLSGM, Harring2021piece}. These parametric and semi-parametric models are effective but may be limited when the true nature of the change is not well understood in advance. Herein lies the benefit of latent basis growth models (LBGMs). These models afford greater flexibility by allowing researchers to determine an optimal curve shape without prior assumptions \citep{McArdle1987LBGM, Meredith1990LBGM}. Our work leverages this flexibility to enable more nuanced data explorations, meeting the demand for more adaptable tools in longitudinal data analysis.

\subsection{Traditional Specification of Latent Basis Growth Model}\label{I:old_LBGM}
\citet[Chapter~11]{Grimm2016growth} demonstrates that LBGMs can be constructed using both the Latent Growth Curve Modeling (LGCM) framework, a subset of the Structural Equation Modeling (SEM) framework, and the mixed-effects modeling framework. While LBGMs were not explicitly discussed, the existing literature suggests that, for a majority of longitudinal models, these two frameworks are mathematically equivalent in evaluating between-individual differences in within-individual changes \citep{Bauer2003equal, Curran2003equal}. This study focuses on LGCM framework due to its greater modeling flexibility and widespread recognition within the social science research community.

Similar to other latent growth curve models, a LBGM can be expressed as $\boldsymbol{y}_{i}=\boldsymbol{\Lambda}\boldsymbol{\eta}_{i}+\boldsymbol{\epsilon}_{i}$, where $\boldsymbol{y}_{i}$ represents the repeated measurements for individual $i$, $\boldsymbol{\eta}_{i}$ are the latent growth factors, $\boldsymbol{\Lambda}$ is the matrix of factor loadings, and $\boldsymbol{\epsilon}_{i}$ is residual vector of individual $i$. Simply put, this equation captures how an individual's growth pattern is influenced by latent factors and measurement occasions. LBGMs usually consist of two growth factors, representing an intercept and a shape factor. 

The factor loading matrix $\boldsymbol{\Lambda}$ is partially constrained for model identification. Specifically, in a setting with $J$ measurements, factor loadings for the intercept are fixed at 1, while two factor loadings for the shape are also fixed, and the remaining $J-2$ are estimated. This arrangement is not arbitrary; it is essential for model identification and provides flexibility in capturing different growth patterns. Figures \ref{fig:LBGM_old1} and \ref{fig:LBGM_old2} illustrate two common specifications of LBGM with six repeated measurements. In Figure \ref{fig:LBGM_old1}, the shape factor is scaled based on the change during the initial time interval. In Figure \ref{fig:LBGM_old2}, the shape factor is scaled based on the total change over the study duration. These methods allow for the flexible estimation of $\boldsymbol{\Lambda}$, thus liberating LBGM from being restricted to a specific functional form. To summarize, the flexibility in specifying $\boldsymbol{\Lambda}$ allows LBGMs to adapt to different research questions and datasets, making them a powerful tool for longitudinal data analysis.

\figurehere{1}

\subsection{Novel Specification of Latent Basis Growth Model}\label{I:new_LBGM}
Although the LBGM described in Section \ref{I:old_LBGM} is a flexible statistical tool to explore trajectories: neither whether nonlinearity exists \citep{Grimm2013LBGM} nor any details of the nature of the nonlinearity need to specify \citep{Wood2015LBGM}, it still has limitations. According to \citet[Chapter~11]{Grimm2016growth}, discrete intervals of time are required when specifying a LBGM, and therefore, it cannot be fit in the framework of individual measurement occasions. We can approximate continuous measurement time using the time-bins approach, also known as the time-windows method. However, several studies highlight the drawbacks of this approach. For example, \citet{Blozis2008coding} have demonstrated that using the time-bins approach may lead to inadmissible estimation, such as overestimating the within-person changes and underestimating the between-person differences, though sometimes these effects are negligible if the degree of individual differences is not notable. Moreover, \citet{Coulombe2015ignoring} concluded that neglecting time differences often leads to undesirable outcomes, such as biased parameter estimates through evaluating the bias of the estimated parameters, efficiency, and Type I error rate with different combinations of sample size, the degree of heterogeneity, the distribution of time, the rate of change, and the number of repeated measurements. 

Two parallel but distinct methods for accounting for individual measurement occasions have been proposed by \citet{Sterba2014individually} and \citet{Liu2022LCSM}. \citet{Sterba2014individually} introduces two growth factors—the intercept and shape factor—and defines shape factor loadings as a function of individual measurement times and their deviations from linearity. In contrast, the framework by \citet{Liu2022LCSM} extends the latent basis growth model by considering linear piecewise functional forms with $J$ measurements and $J-1$ segments. Building upon the conceptual framework introduced by \citet{Liu2022LCSM}, one can examine both the growth status and growth rate of a LBGM, as depicted in Figures \ref{fig:LBGM_growth} and \ref{fig:LBGM_rate}, respectively. Specifically, Figure \ref{fig:LBGM_growth} illustrates the change in growth status, highlighted by a difference of $0.8$ calculated for the interval from $t=1$ to $t=2$. This change can also be quantified using the area under the curve (AUC) in Figure \ref{fig:LBGM_rate}, which effectively represents the integral of the growth rate over that time interval. As an example, $0.8 \times (2-1)=0.8$ for the interval from $t=1$ to $t=2$. Extending from this concept of utilizing the AUC to represent change, \citet{Liu2022LCSM} proposed a novel specification for LBGMs. Figure \ref{fig:LBGM_new1} demonstrates the path diagram of this new specification for a LBGM with six measurement occasions. Here, two growth factors, $\eta_{0}$ and $\eta_{1}$, define the intercept and first-interval growth rate for each individual. With this specification in place, factor loadings can be expressed as the product of the relative growth rate $\gamma_{j-1}$, defined as interval-specific slopes divided by $\eta_{1}$, and the corresponding time intervals. These time intervals are allowed to be individually different, which are `definition variables' according to \citet{Mehta2000people, Mehta2005people}, and \citet{Sterba2014individually}.

\figurehere{2}

\figurehere{3}

In addition to the standard specifications of LBGM, this framework provides flexibility in scaling the growth rate factor, $\eta_{1}$. Contrary to constraining $\eta_{1}$ to the first time interval, it can be adapted to represent growth during any selected time frame, for example, the last time interval as demonstrated in Figure \ref{fig:LBGM_new2}. Here, $\gamma_{j-1}$ continues to serve as the relative growth rate in relation to $\eta_{1}$ for each $(j-1)^{th}$ time interval. With this novel approach, the shape factor's loading at each measurement occasion $t_{j}$ is determined by dividing the change-from-baseline (the difference between the current value and the initial value at $t_1$) at $t_{j}$ by $\eta_{1}$.

\subsection{Parallel Latent Basis Growth Model}\label{I:PLCSM}
In the study of joint longitudinal processes, researchers commonly employ Multivariate Growth Models (MGMs) \citep[Chapter~8]{Grimm2016growth}, also known as parallel process and correlated growth models \citep{McArdle1988Multi}. MGMs can be categorized into three main types of associations: (1) within-construct growth factors, (2) between-construct growth factors, and (3) between-construct residuals \citep[Chapter~8]{Grimm2016growth}. Existing research has contributed valuable frameworks for understanding these relationships. For example, \citet{Robitaille2012PLGM} employed a MGM with linear growth curves to investigate the co-evolution of processing speed and visuospatial ability. Similarly, \citet{Blozis2004MGM, Blozis2008MGM} introduced MGMs that accommodate parametric nonlinear functional forms such as polynomial and exponential curves. More recently, \citet{Peralta2020PBLSGM} and \citet{Liu2021PBLSGM} developed parallel bilinear spline growth models with unknown random knots, leveraging Bayesian mixed-effects and frequentist structural equation frameworks, respectively. While these models have proven to be effective for theory-driven inquiries, they may lack the flexibility required during the exploratory stages of research, particularly when there is no guiding domain-specific theory for model selection.

This article aims to advance the field of joint longitudinal process modeling by extending the LBGM with a novel specification, detailed in Section \ref{I:new_LBGM}. Notably, our extension adapts the LBGM to function within the framework of individual measurement occasions. This adaptation enables the model to account for different numbers of measurements or varied sets of measurement occasions across constructs, thus providing a more versatile tool for analyzing longitudinal data.

The proposed model fills existing gaps by demonstrating how to fit a parallel LBGM in the framework of individual measurement occasions. The remainder of this article is organized as follows. We start from a latent basis growth model for a univariate longitudinal process with the novel specification in the method section. We then extend it to a parallel growth curve model and introduce the model specification and estimation of the parallel model. Next, we describe the model evaluation that is realized by the Monte Carlo simulation study. We evaluate how the proposed model works through performance measures, including the relative bias, the empirical standard error (SE), the relative root-mean-square-error (RMSE), and the coverage probability (CP) of a $95\%$ confidence interval. We also demonstrate how to apply the proposed model by analyzing a real-world data set of longitudinal reading scores and mathematics scores from Early Childhood Longitudinal Study, Kindergarten Class of $2010$-$11$ (ECLS-K: $2011$). In this application section, we also show how to interpret possible insights from the model output. Finally, discussions are framed regarding practical considerations, methodological considerations, and future directions.

\section{Method}\label{sec:Method}
\subsection{Latent Basis Growth Model in the Framework of Individual Measurement Occasions}\label{M:Univariate}
This section introduced the novel specification of a LBGM developed in \citet{Liu2022LCSM} for univariate nonlinear developmental trajectories. This model can, for instance, be applied to analyze univariate longitudinal outcome such as reading or mathematics development. It can be expressed as
\begin{align}
&\boldsymbol{y}_{i} = \boldsymbol{\Lambda}^{[y]}_{i} \times \boldsymbol{\eta}^{[y]}_{i} + \boldsymbol{\epsilon}^{[y]}_{i}, \label{eq:uni_LBGM}, \\
&\boldsymbol{\eta}^{[y]}_{i} = \boldsymbol{\mu}^{[y]}_{\boldsymbol{\eta}} + \boldsymbol{\zeta}^{[y]}_{i}, \label{eq:uni_LBGM_gf}
\end{align}
where $\boldsymbol{y}_{i}$ is a $J \times 1$ vector representing the $i^{th}$ individual's repeated measurements (with $J$ denoting the number of such measurements). The vector $\boldsymbol{\eta}^{[y]}_{i}$ is a $2 \times 1$ vector of growth factors, where the first element ($\eta_{0i}$) signifies the initial status and the second element ($\eta_{1i}$) indicates the growth rate within a specified time interval. The $J \times 2$ matrix $\boldsymbol{\Lambda}^{[y]}_{i}$ consists of associated factor loadings. Finally, $\boldsymbol{\epsilon}^{[y]}_{i}$ is a $J \times 1$ vector of the $i^{th}$ individual's residuals. Equation \eqref{eq:uni_LBGM_gf} expresses $\boldsymbol{\eta}^{[y]}_{i}$ as deviations ($\boldsymbol{\zeta}^{[y]}_{i}$) from the mean values of the growth factors ($\boldsymbol{\mu}^{[y]}_{\boldsymbol{\eta}}$).

In \citet{Liu2022LCSM}, the term $\eta_{1i}$ is scaled to represent the growth rate during the first time interval, as illustrated in Figure \ref{fig:LBGM_new1}. However, this term can also be scaled to correspond with the growth rate during any other time interval, such as the last one depicted in Figure \ref{fig:LBGM_new2}. Although the choice of scaling affects the interpretation of $\eta_{1i}$, the general form of the matrix of factor loadings, $\boldsymbol{\Lambda}^{[y]}_{i}$, remains consistent. This general form is given as:
\begin{equation}\label{eq:uni_loadings}
\boldsymbol{\Lambda}^{[y]}_{i}=\begin{pmatrix}
1 & 0 \\
1 & \gamma^{[y]}_{2-1}\times(t_{i2}-t_{i1}) \\
1 & \sum_{j=2}^{3}\gamma^{[y]}_{j-1}\times(t_{ij}-t_{i(j-1)}) \\
\dots & \dots \\
1 & \sum_{j=2}^{J}\gamma^{[y]}_{j-1}\times(t_{ij}-t_{i(j-1)}) \\
\end{pmatrix},
\end{equation}
where the $j^{th}$ element of the second column represents the change-from-baseline divided by $\eta_{1i}$ at the corresponding measurement occasion of the $i^{th}$ individual. In Equation (\ref{eq:uni_loadings}), $\gamma_{j-1}$ $(j=2, \dots, J)$ is the relative growth rate to $\eta_{1i}$ during the $(j-1)^{th}$ time interval. To specify the scaling of $\eta_{1i}$, one fixes the corresponding $\gamma_{j-1}$ to $1$. For instance, \citet{Liu2022LCSM} fixed $\gamma_{2-1}$ to $1$ to indicate that $\eta_{1i}$ is scaled to the first time interval, as shown in Figure \ref{fig:LBGM_new1}. Similarly, to scale $\eta_{1i}$ to the last time interval, one would fix $\gamma_{J-1}$ to $1$, as shown in Figure \ref{fig:LBGM_new2}. It is worth noting that the subscript $i$ in $\boldsymbol{\Lambda}^{[y]}_{i}$ underlines that the model operates within the framework of individual measurement occasions.

\subsection{Model Specification of Parallel Latent Basis Growth Model}\label{M:Parallel}
In this section, we extend the univariate Latent Basis Growth Model (LBGM) to its parallel version. This parallel version enables the joint analysis of multiple repeated outcomes, such as the joint development of reading and mathematics ability. The necessity for this extension arises from the various compelling reasons that have been discussed in Section \ref{sec:Intro}. We describe the parallel LBGM in the context of individual measurement occasions, extending the univariate model given in Equation (\ref{eq:uni_LBGM}). Assume that we have bivariate growth trajectories for repeated outcomes, the parallel LBGM can then be formally defined as follows:
\begin{equation}\label{eq:bi_LBGM}
\begin{pmatrix}
\boldsymbol{y}_{i} \\ \boldsymbol{z}_{i}
\end{pmatrix}=
\begin{pmatrix}
\boldsymbol{\Lambda}_{i}^{[y]} & \boldsymbol{0} \\ \boldsymbol{0} & \boldsymbol{\Lambda}_{i}^{[z]}
\end{pmatrix}\times
\begin{pmatrix}
\boldsymbol{\eta}^{[y]}_{i} \\ \boldsymbol{\eta}^{[z]}_{i}
\end{pmatrix}+
\begin{pmatrix}
\boldsymbol{\epsilon}^{[y]}_{i} \\ \boldsymbol{\epsilon}^{[z]}_{i}
\end{pmatrix},
\end{equation}
where $\boldsymbol{z}_{i}$ is also a $J\times 1$ vector of the repeated measurements for individual $i$, $\boldsymbol{\eta}^{[z]}_{i}$, $\boldsymbol{\Lambda}_{i}^{[z]}$ and $\boldsymbol{\epsilon}^{[z]}_{i}$ are its growth factors (a $2\times1$ vector), the corresponding factor loadings (a $J\times2$ matrix), and the residuals of person $i$ (a $J\times 1$ vector), respectively. Similar to $\boldsymbol{\Lambda}_{i}^{[y]}$, $\boldsymbol{\Lambda}_{i}^{[z]}$ has a general expression but with one fixed relative growth rate $\gamma_{j-1}$, corresponding to the growth rate of $(j-1)^{th}$ time interval that $\eta^{[z]}_{1i}$ represents. We then write the outcome-specific growth factors $\boldsymbol{\eta}^{[u]}_{i}$ ($u=y,\ z$) as deviations from the corresponding outcome-specific growth factor means.
\begin{equation}\label{eq:bi_LBGM_gf}
\begin{pmatrix}
\boldsymbol{\eta}^{[y]}_{i} \\ \boldsymbol{\eta}^{[z]}_{i}
\end{pmatrix}=
\begin{pmatrix}
\boldsymbol{\mu}^{[y]}_{\boldsymbol{\eta}} \\ \boldsymbol{\mu}^{[z]}_{\boldsymbol{\eta}}
\end{pmatrix}+
\begin{pmatrix}
\boldsymbol{\zeta}^{[y]}_{i} \\
\boldsymbol{\zeta}^{[z]}_{i}
\end{pmatrix},
\end{equation}
where $\boldsymbol{\mu}^{[u]}_{\boldsymbol{\eta}}$ is a $2\times 1$ vector of outcome-specific growth factor means, and $\boldsymbol{\zeta}^{[u]}_{i}$ is a $2\times 1$ vector of deviations of the $i^{th}$ individual from the means. To simplify model, we assume that $\begin{pmatrix} \boldsymbol{\zeta}^{[y]}_{i} & \boldsymbol{\zeta}^{[z]}_{i}\end{pmatrix}^{T}$ follows a multivariate normal distribution
\begin{equation}\nonumber
\begin{pmatrix} 
\boldsymbol{\zeta}^{[y]}_{i} \\ \boldsymbol{\zeta}^{[z]}_{i}
\end{pmatrix}\sim \text{MVN}\bigg(\boldsymbol{0}, 
\begin{pmatrix}
\boldsymbol{\Psi}_{\boldsymbol{\eta}}^{[y]} & \boldsymbol{\Psi}_{\boldsymbol{\eta}}^{[yz]} \\
& \boldsymbol{\Psi}_{\boldsymbol{\eta}}^{[z]}
\end{pmatrix}\bigg),
\end{equation}
where both $\boldsymbol{\Psi}_{\boldsymbol{\eta}}^{[u]}$ and $\boldsymbol{\Psi}_{\boldsymbol{\eta}}^{[yz]}$ are $2\times 2$ matrices: $\boldsymbol{\Psi}_{\boldsymbol{\eta}}^{[u]}$ is the variance-covariance matrix of the outcome-specific growth factors while $\boldsymbol{\Psi}_{\boldsymbol{\eta}}^{[yz]}$ is the covariances between the growth factors of $\boldsymbol{y}_{i}$ and $\boldsymbol{z}_{i}$. To simplify the model, we also assume that the individual outcome-specific residual variances are identical and independent normal distributions over time, while the residual covariances are homogeneous over time, that is,
\begin{equation}\nonumber
\begin{pmatrix} 
\boldsymbol{\epsilon}^{[y]}_{i} \\ \boldsymbol{\epsilon}^{[z]}_{i}
\end{pmatrix}\sim \text{MVN}\bigg(\boldsymbol{0}, 
\begin{pmatrix}
\theta^{[y]}_{\epsilon}\boldsymbol{I} & \theta^{[yz]}_{\epsilon}\boldsymbol{I} \\
& \theta^{[z]}_{\epsilon}\boldsymbol{I}
\end{pmatrix}\bigg),
\end{equation}
where $\boldsymbol{I}$ is a $J\times J$ identity matrix.

\subsection{Model Estimation}\label{M:estimate}
We then write the expected mean vector and variance-covariance matrix of the bivariate repeated outcome $\boldsymbol{y}_{i}$ and $\boldsymbol{z}_{i}$ in the parallel LBGM specified in Equations (\ref{eq:bi_LBGM}) and (\ref{eq:bi_LBGM_gf}) as
\begin{equation}\label{eq:mean_r}
\boldsymbol{\mu}_{i}=\begin{pmatrix}
\boldsymbol{\mu}^{[y]}_{i} \\ \boldsymbol{\mu}^{[z]}_{i}
\end{pmatrix}=\begin{pmatrix}
\boldsymbol{\Lambda}_{i}^{[y]} & \boldsymbol{0} \\ \boldsymbol{0} & \boldsymbol{\Lambda}_{i}^{[z]}
\end{pmatrix}\times\begin{pmatrix}
\boldsymbol{\mu}^{[y]}_{\boldsymbol{\eta}} \\ \boldsymbol{\mu}^{[z]}_{\boldsymbol{\eta}}
\end{pmatrix}
\end{equation}
and
\begin{equation}\label{eq:var_r}
\begin{aligned}
\boldsymbol{\Sigma}_{i}&=\begin{pmatrix}
\boldsymbol{\Sigma}^{[y]}_{i} & \boldsymbol{\Sigma}^{[yz]}_{i} \\
& \boldsymbol{\Sigma}^{[z]}_{i}
\end{pmatrix}\\
&=\begin{pmatrix}
\boldsymbol{\Lambda}_{i}^{[y]} & \boldsymbol{0} \\ \boldsymbol{0} & \boldsymbol{\Lambda}_{i}^{[z]}
\end{pmatrix}\times\begin{pmatrix}
\boldsymbol{\Psi}^{[y]}_{\boldsymbol{\eta}} & \boldsymbol{\Psi}^{[yz]}_{\boldsymbol{\eta}} \\
& \boldsymbol{\Psi}^{[z]}_{\boldsymbol{\eta}}
\end{pmatrix}\times\begin{pmatrix}
\boldsymbol{\Lambda}_{i}^{[y]} & \boldsymbol{0} \\ \boldsymbol{0} & \boldsymbol{\Lambda}_{i}^{[z]}
\end{pmatrix}^{T}\\
&+\begin{pmatrix}
\theta^{[y]}_{\epsilon}\boldsymbol{I} & \theta^{[yz]}_{\epsilon}\boldsymbol{I} \\
& \theta^{[z]}_{\epsilon}\boldsymbol{I}
\end{pmatrix}.
\end{aligned}
\end{equation}

The parameters in the parallel LBGM specified in Equations (\ref{eq:bi_LBGM}) and (\ref{eq:bi_LBGM_gf}) include the mean vector and variance-covariance matrix of the growth factors, the outcome-specific relative growth rate, the variance-covariance matrix of the residuals. Accordingly, we define
\begin{equation}\label{eq:theta}
\begin{aligned}
\boldsymbol{\Theta}=&\{\boldsymbol{\mu}^{[u]}_{\boldsymbol{\eta}}, \boldsymbol{\Psi}^{[u]}_{\boldsymbol{\eta}}, \boldsymbol{\Psi}^{[yz]}_{\boldsymbol{\eta}}, 
\boldsymbol{\gamma}^{[u]}, \theta^{[u]}_{\epsilon}, \theta^{[yz]}_{\epsilon}\}\\
=&\{\mu^{[u]}_{\eta_{0}}, \mu^{[u]}_{\eta_{1}}, \psi^{[u]}_{00}, \psi^{[u]}_{01}, \psi^{[u]}_{11}, \psi^{[yz]}_{00}, \psi^{[yz]}_{01}, \psi^{[yz]}_{10}, \psi^{[yz]}_{11}, \gamma^{[u]}_{j-1}, \\
&\theta^{[u]}_{\epsilon}, \theta^{[yz]}_{\epsilon}\},\ u=y,\ z\\
&j=\begin{cases}
3, \dots, J & \text{Model specification in Figure \ref{fig:LBGM_new1}} \\
2, \dots, J-1 & \text{Model specification in Figure \ref{fig:LBGM_new2}}
\end{cases}
\end{aligned}
\end{equation}
to list the parameters that we need to estimated in the proposed model. 

We estimate $\boldsymbol{\Theta}$ using full information maximum likelihood (FIML) to account for the individual measurement occasions and potential heterogeneity of individual contributions to the likelihood function. In this present study, the proposed model is built using the R package \textit{OpenMx} with CSOLNP optimizer \citep{Pritikin2015OpenMx, OpenMx2016package, User2020OpenMx, Hunter2018OpenMx}. We provide \textit{OpenMx} code of the proposed parallel LBGM and a demonstration in the online appendix (\url{https://github.com/#####/Extension_projects}). We also provide \textit{Mplus} 8 code of the proposed model for researchers who are interested in using \textit{Mplus}.

\section{Model Evaluation}\label{M:Evaluate}
We aim to assess the effectiveness of the proposed parallel LBGM by employing Monte Carlo simulation studies. Specifically, we examine the model's performance using several metrics: the relative bias, the empirical standard error (SE), the relative root-mean-square error (RMSE), and the empirical coverage probability for a nominal $95\%$ confidence interval for each parameter. These metrics are commonly used in simulation studies to evaluate the performance of statistical methodologies or models. The definitions and estimates for these metrics are presented in Table \ref{tbl:metric}.

\tablehere{1}

Following practices in simulation studies as suggested by \citet{Morris2019simulation}, we empirically determined the number of replications to be $S=1,000$. Among the four performance metrics, the (relative) bias is of utmost importance. A pilot simulation showed that the standard errors of bias, calculated as $\text{Monte Carlo SE(Bias)} = \sqrt{\frac{\text{Var}(\hat{\theta})}{S}}$, were less than $0.15$ across all parameters, except for $\psi_{00}^{[u]}$ and $\psi_{00}^{[yz]}$. To maintain the Monte Carlo standard error of bias below $0.05$, at least $900$ replications are needed. We decided to proceed with $S=1,000$ replications to account for variability and ensure a more robust evaluation.

\subsection{Design of Simulation Study}\label{Evaluation:design}
In order to thoroughly evaluate the proposed parallel LBGM, we designed a comprehensive set of simulation studies, the conditions of which are outlined in Table \ref{tbl:simu_design}. The primary factor in the effectiveness of a model designed for longitudinal data is the number of repeated measures. We hypothesize that the proposed model's performance would improve with an increasing number of repeated measurements. To test this hypothesis, we considered two levels for the number of repeated measures: six and ten. For conditions with ten repeated measures, we aim to investigate whether equally-placed study waves or not affects model performance, assuming that the study duration is constant across conditions. In the scenario with six repeated measures, we examine the model's performance under a more challenging condition with a shorter study duration. Measurement occasions are individuated by a `medium'-width time window, $(-0.25, +0.25)$ around each wave \citep{Coulombe2015ignoring}.

\tablehere{2}

Another key variable of interest is the correlation between the two trajectories, as the proposed model is designed for analyzing joint longitudinal processes. Three correlation levels for the between-construct growth factors are considered: $\pm 0.3$ and $0$. We are interested in how model over-specification affects performance in zero-correlation conditions, and whether the sign of the correlation ($\pm 0.3$) has any impact on model performance. Additionally, we explore the influence of varying trajectory shapes, quantified by the relative growth rate in each time interval. As specified in Table \ref{tbl:simu_design}, the change patterns considered include both increasing and decreasing growth rates. Moreover, we evaluate the model's performance across different sample sizes ($n=200$ and $n=500$) and levels of outcome-specific residual variances ($\theta^{[u]}_{\epsilon}=1$or $\theta^{[u]}_{\epsilon}=2$) to gauge the effects of sample size and measurement precision. In the simulation design, factors deemed non-influential to the proposed model's performance, such as the distribution of growth factors and the correlation of between-construct residuals, were held constant.

\subsection{Data Generation and Simulation Step}\label{evaluation:step}
To evaluate the performance of the proposed parallel LBGMs, we conducted a simulation study according to the design presented in Table \ref{tbl:simu_design}. Each condition was replicated $1,000$ times to ensure a robust assessment. The steps for the simulation are outlined as follows:
\begin{enumerate}
\item \textbf{Growth Factor Generation:} Utilizing the \textit{MASS} R package \citep{Venables2002Statistics}, generate the growth factors for both longitudinal processes based on the pre-defined mean vector and variance-covariance matrix as specified in Table \ref{tbl:simu_design}. The \textit{MASS} package is used for its reliability in generating multivariate Gaussian samples.
\item \textbf{Time Structure:} Generate the time structure with $J$ waves $t_{j}$ as defined in Table \ref{tbl:simu_design}. Add a uniform disturbance following $U(t_{j} - \Delta, t_{j} + \Delta)$ around each wave to obtain individual measurement occasions $t_{ij}$.
\item \textbf{Factor Loadings Calculation:} Compute the factor loadings for each individual of each construct, using the relative growth rates and individual measurement intervals.
\item \textbf{Measurement Value Computation:} Calculate the values of bivariate repeated measurements, incorporating growth factors, factor loadings, and the pre-defined residual variance-covariance structure.
\item \textbf{LBGM Implementation:} Execute the proposed LBGM model on the generated dataset, estimating the model parameters and constructing $95\%$ Wald confidence intervals.
\item \textbf{Replication:} Repeat steps 1--5 until $1,000$ convergent solutions are obtained, as this number of replications provides a stable estimate of performance metrics such as bias and coverage probability.
\end{enumerate}

\section{Result}\label{sec:Result}
\subsection{Model Convergence}\label{R:Preliminary}
Before assessing the four performance measures of the proposed parallel LBGM, we first examined its convergence rate\footnote{In this study, we define convergence rate as the achievement of an \textit{OpenMx} status code of $0$, indicating successful optimization, in up to $10$ runs with varied initial values \citep{OpenMx2016package}.}. The model exhibited excellent convergence, as evidenced by a $100\%$ rate across all simulation conditions listed in Table \ref{tbl:simu_design}.

\subsection{Performance Measures}\label{R:Primary}
This section summarizes the simulation results for four key performance metrics: relative bias, empirical SE, relative RMSE, and empirical coverage probability for a nominal $95\%$ confidence interval. We calculated these metrics for each parameter across $1,000$ repetitions under each condition, and summarized the median and range values for all conditions given the scale of parameters and simulation setups. The proposed model generally yielded unbiased and accurate point estimates with target coverage probabilities. Further details for each performance metric are provided in the Online Supplementary Document.

The proposed model produced unbiased point estimates and low empirical SEs. Specifically, the magnitudes of the relative biases for outcome-specific growth factor means, variances, and relative growth rates were below $0.004$, $0.013$, and $0.012$, respectively. The empirical SE magnitudes for all parameters, except for $\psi_{00}^{[u]}$ and $\psi_{00}^{[yz]}$, were below $0.45$. The model also generated accurate estimates; the magnitudes of the relative RMSEs for outcome-specific growth factor means, variances, and relative growth rates were below $0.05$, $0.15$, and $0.23$, respectively. Moreover, the model demonstrated excellent empirical coverage probabilities, with median values approximating $0.95$. Given these consistently strong performance metrics, further investigations into the effect of different simulation conditions were deemed unnecessary.

\section{Application}\label{Sec:Application}
In this section, we demonstrate how to employ the proposed parallel LBGM to analyze real-world data. This application section includes two examples. In the first example, we illustrate the recommended steps to construct the proposed model in practice. In the second example, we demonstrate how to apply the proposed model to analyze joint longitudinal processes with a more complicated data structure where two repeated outcomes have different time frames. We randomly selected $400$ students from the Early Childhood Longitudinal Study Kindergarten Cohort of 2010-2011 (ECLS-K: 2011), all of whom had complete records of repeated reading and mathematics scores based on Item Response Theory (IRT), as well as their age in months at each wave\footnote{The total sample size of ECLS-K: 2011 is $n=18174$. The number of rows after removing records with missing values (i.e., entries with any of NaN/-9/-8/-7/-1) is $n=2290$.}.

ECLS-K: 2011 is a national longitudinal study of US children registered in around $900$ kindergarten programs beginning in $2010$-$2011$ school year. In ECLS-K: 2011, children's reading ability and mathematics ability were assessed in nine waves: fall and spring of kindergarten ($2010$-$2011$), first ($2011$-$2012$) and second ($2012$-$2013$) grade, respectively, as well as spring of the $3^{rd}$ ($2014$), $4^{th}$ ($2015$) and $5^{th}$ ($2016$) grade, respectively. Only about $30\%$ of students were assessed in the fall semesters of 2011 and 2012. \citep{Le2011ECLS}. In the first example, we used all nine waves of reading and mathematics IRT scores to demonstrate how to apply the proposed model. In the second example, we utilized all nine waves of reading IRT scores but only the mathematics scores obtained in spring semesters to mimic one possible complex time structure in practice. Note that the initial status and the number of measurement occasions of the two abilities are different in the second example. Additionally, we employed children's age in months rather than their grade-in-school to have individual measurement occasions. The subsample included $41.50\%$ White, $7.25\%$ Black, $37.00\%$ Latinx, $8.25\%$ Asian, and $6.00\%$ of other ethnicity.

\subsection{Analyze Joint Longitudinal Records with The Same Time Structure}
Following \citet{Blozis2008MGM, Liu2021PBLSGM}, we first built a latent growth curve model to examine each longitudinal process in isolation before analyzing joint development. Specifically, we employed a LBGM to explore the univariate development of either reading or mathematics from Grade K to 5. Figure \ref{fig:Uni_traj} illustrates the model-implied curves superimposed on the smooth lines for each ability. For each ability, the estimates from the LBGM produced model-implied trajectories that closely align with the smooth lines representing the observed individual data.

\figurehere{4}

We then applied the proposed parallel LBGM to analyze the joint development of reading and mathematics ability. Figure \ref{fig:Bi_traj} demonstrates the model implied curves on the smooth lines for each ability obtained from the parallel model. From the figure, we can see that the model implied curves of parallel models did not change from those of univariate growth models shown in Figure \ref{fig:Uni_traj}. Table \ref{tbl:RM_est} presents the parameter estimates of interest for joint development.

\figurehere{5}

\tablehere{3}

Note that we defined $\eta^{[u]}{1}$ as the growth rate in the final time interval for each ability's longitudinal process (i.e., the model specification in Figure \ref{fig:LBGM_new2}). That is, in Table \ref{tbl:RM_est}, the parameters related to `initial status' and `rate of Interval $8$' were estimated from the proposed model directly, while others were obtained by the function \textit{mxAlgebra()}\footnote{By using \textit{mxAlgebra()}, we need to specify algebraic expression of new parameters, then \textit{OpenMx} is capable of their point estimates along with standard errors.} in the \textit{R} package \textit{OpenMx}. From Figure \ref{fig:Bi_traj} and Table \ref{tbl:RM_est}, we observed that the development of both reading and mathematics ability slowed down post-Grade $3$ in general, which aligns with earlier studies \citep{Peralta2020PBLSGM, Liu2021PBLSGM}. In addition, there was a positive association between the development of reading and mathematics ability indicated by statistically significant intercept-intercept and slope-slope covariance in each time interval. 

Standardizing the covariances, the intercept-intercept correlation and each interval-specific slope-slope correlation were $0.83$ and $0.58$, respectively. Therefore, it suggests that, on average, a child who performed better in reading tests at Grade K tended to perform better in mathematics examinations and vice versa. Moreover, on average, children who showed more rapid gains in reading ability also tended to exhibit faster improvement in mathematics, and vice versa.

\subsection{Analyze Joint Longitudinal Records with Different Time Structures}
In this section, we use the proposed parallel LBGM to investigate the joint development trajectories of reading and mathematics abilities, where we kept all nine measurement occasions of reading ability but only the measurements of mathematics ability in the spring semesters (i.e., Wave $2$, $4$, $6$, $7$, $8$, and $9$). In this configuration, both the initial statuses and the numbers of measurement occasions differ between the two abilities. Figure \ref{fig:Bi_traj_diff} illustrates the model-implied curves superimposed on the smooth lines representing each ability in this model. The figure reveals that the model-implied trajectories vary only minimally from those presented in Figure \ref{fig:Bi_traj} due to fewer measurement occasions in mathematics ability, but it still sufficiently captured the smooth lines of observed individual data. 

\figurehere{6}

Table \ref{tbl:RM_est_diff} presents the estimated parameters of interest for the joint model with differing time structures. Note that we have $8$ time intervals of the development of the reading ability (corresponding to $9$ measurement occasions) but only $5$ time intervals of the development of mathematics ability because we took out three measurements in fall semesters. During the first time interval for mathematics, corresponding to Intervals $2$ and $3$ for reading ability (as detailed in Table \ref{tbl:RM_est}), the estimated growth rate was $1.811$. This is an average of the growth rates $1.437$ and $2.169$ from Interval $2$ and Interval $3$, respectively, in Table \ref{tbl:RM_est}. These findings suggest that our proposed model effectively captures the underlying patterns of growth trajectories, even with fewer measurements.

\tablehere{4}

\section{Discussion}\label{sec:Discussion}
This article extends the latent basis growth model to explore joint nonlinear longitudinal processes in the framework of individual measurement occasions. This framework is particularly advantageous when investigating parallel development because it helps avoid inadmissible estimation and allows for different time structures across outcomes. Additionally, the proposed model allows scaling the second growth factor as the growth rate during any time interval. In the present study, we specify the second growth factor as the growth rate during either the first or last time interval and estimate the relative rates for each of the other intervals for each repeated outcome. We demonstrate that the proposed parallel LBGM can provide unbiased and accurate point estimates with target coverage probabilities through simulation studies. Additionally, we apply the proposed model to analyze the joint development of reading and mathematics abilities, using the same or different time structures. Our analysis relies on a subsample of $n=400$ from ECLS-K: 
2011.

\subsection{Practical Considerations}\label{D:practical}
In this section, we provide recommendations for empirical researchers based on both the simulation study and real-world data analyses. First, although we scale the shape factor $\eta_{1}$ as the growth rate in the first or last time interval of the study duration, it can be specified as the growth rate in any time interval. Note that the interpretation of $\gamma_{j-1}$ remains as the relative growth rate to $\eta_{1}$ during the $(j-1)^{th}$ time interval. From the proposed parallel LBGM, we obtain the estimates of the mean and variance of shape factor and the fixed effects of relative growth rates for each construct. Using the \textit{mxAlgebra()} function from the \textit{OpenMx} R package, we derive both fixed and random effects for the absolute growth rate of each time interval, as detailed in the Application section.

In addition, the proposed model is capable of estimating the covariance of between-construct intercepts and that of between-construct shape factors directly. We can derive the covariance of between-construct growth rates for each interval by using the function \textit{mxAlgebra()}. Note that the correlation of the between-construct growth rates is constant because we only estimate fixed effects of relative growth rates. 

Third, as the latent basis growth model serves primarily as an exploratory tool, allowing trajectory characteristics to emerge from the data rather than being specified \textit{a priori}, researchers may also be interested in exploring other aspects, such as the change-from-baseline values at each measurement wave for each repeated outcome. We can also derive these features with the function \textit{mxAlgebra()}. In the online appendix (\url{https://github.com/####/Extension_projects}), we also provide code to demonstrate how to derive the values of change-from-baseline. 

\subsection{Methodological Considerations and Future Directions}\label{D:method}
There are several directions to consider for future studies. First, similar to the standard implementation of latent basis growth models, the proposed model requires a strict proportionality assumption \citep{Wu2016LBGM, McNeish2020LBGM}. \citet{Wu2016LBGM} showed that this assumption might potentially result in biased estimates by simulation studies. \citet{McNeish2020LBGM} demonstrated that this assumption could be relaxed by specifying random factor loadings of the shape factor. In the same way, we can also relax the proportionality assumption for the proposed parallel LBGM. Note that the extended model, where both the shape factor and relative growth rates are random coefficients, cannot be specified in a frequentist SEM software because these random coefficients enter the model in a multiplicative fashion (i.e., a nonlinear fashion). Similar to \citet{McNeish2020LBGM}, the extended model can be constructed in Bayesian software such as \textit{jags} or \textit{stan}.

Second, it is not our intention to show that the proposed parallel LBGM is better than any other parallel growth models with parametric or semi-parametric functional forms. The proposed model is a versatile tool for exploratory analyses; it should perform well to detect the trends of trajectories or whether a spike exists over the study duration. However, the insights directly related to research questions might be limited. Accordingly, subsequent analyses may need to be based on the estimates generated by the proposed model. For instance, if we obtain evidence suggesting that developmental processes can generally be divided into two stages, we may employ the parallel bilinear spline growth model \citep{Liu2021PBLSGM} to further estimate the individual transition time to the stage with a slower growth rate. Alternatively, we can constrain the relative growth rates of multiple time intervals to be the same to have a more parsimonious model. Therefore, statistical methods for comparing the full model to a more parsimonious one need to be proposed and tested.

Third, as in any latent growth curve model, baseline covariates can be added to predict the intercept or the growth rate. Additionally, a time-varying covariate can also be added to estimate its effect on the measurements while simultaneously modeling parallel change patterns in these measurements. 

\subsection{Concluding Remarks}\label{D:conclude}
In this article, we propose a novel expression of latent basis growth models to allow for individual measurement occasions and further extend the model to analyze joint longitudinal processes. The results of both the simulation studies and real-world data analyses underscore the model's valuable capabilities for exploring parallel nonlinear change patterns. As discussed above, the proposed method offers avenues for both practical extensions and further methodological examination.


\newpage

\bibliographystyle{apalike}
\bibliography{Extension5}

\renewcommand\thefigure{\arabic{figure}}
\setcounter{figure}{0}
\begin{figure}[!ht]
\centering
\begin{subfigure}{.5\textwidth}
 \centering
 \includegraphics[width=0.95\linewidth]{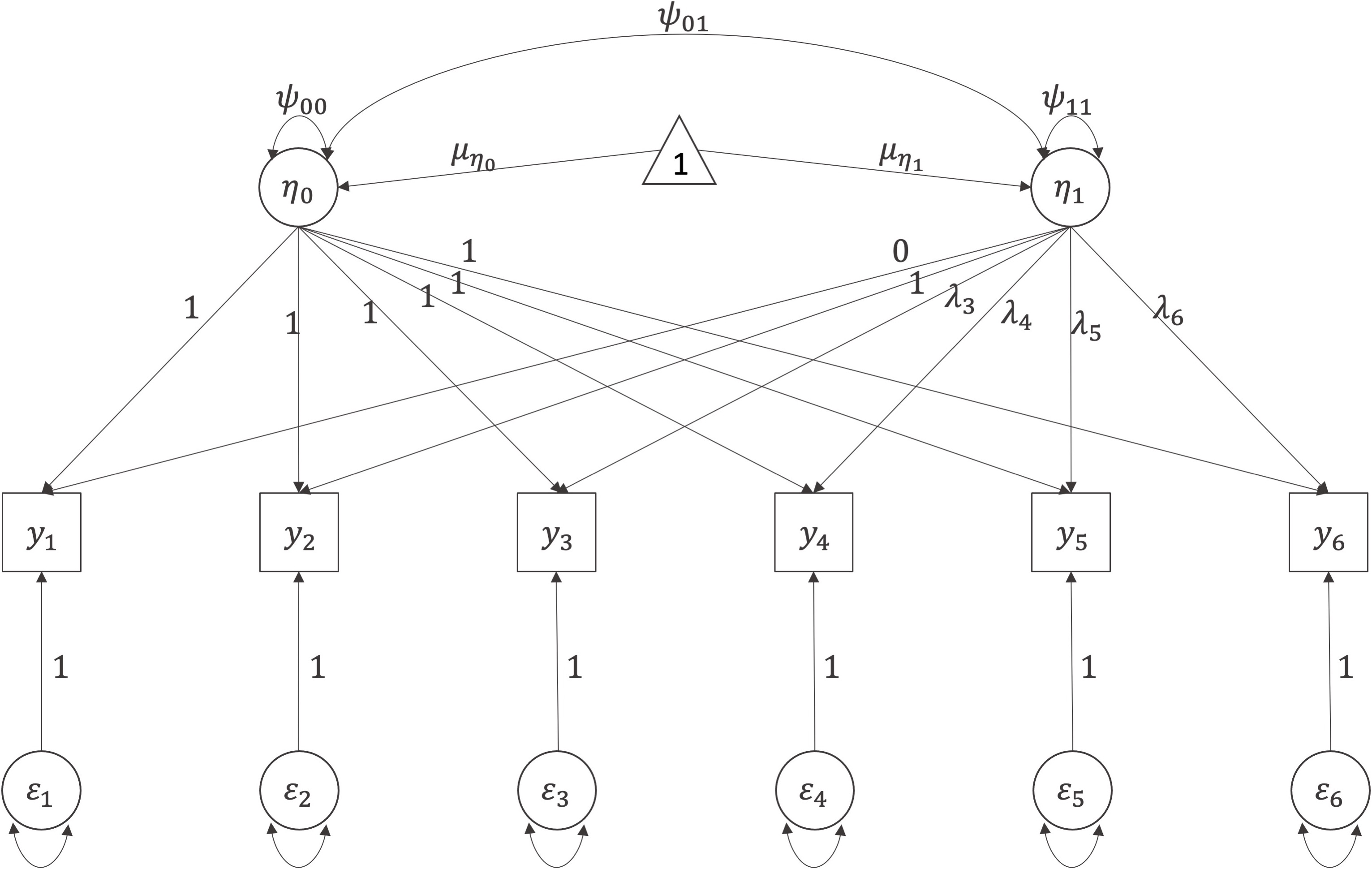}
 \caption{Specification 1}
 \label{fig:LBGM_old1}
\end{subfigure}%
\begin{subfigure}{.5\textwidth}
 \centering
 \includegraphics[width=0.95\linewidth]{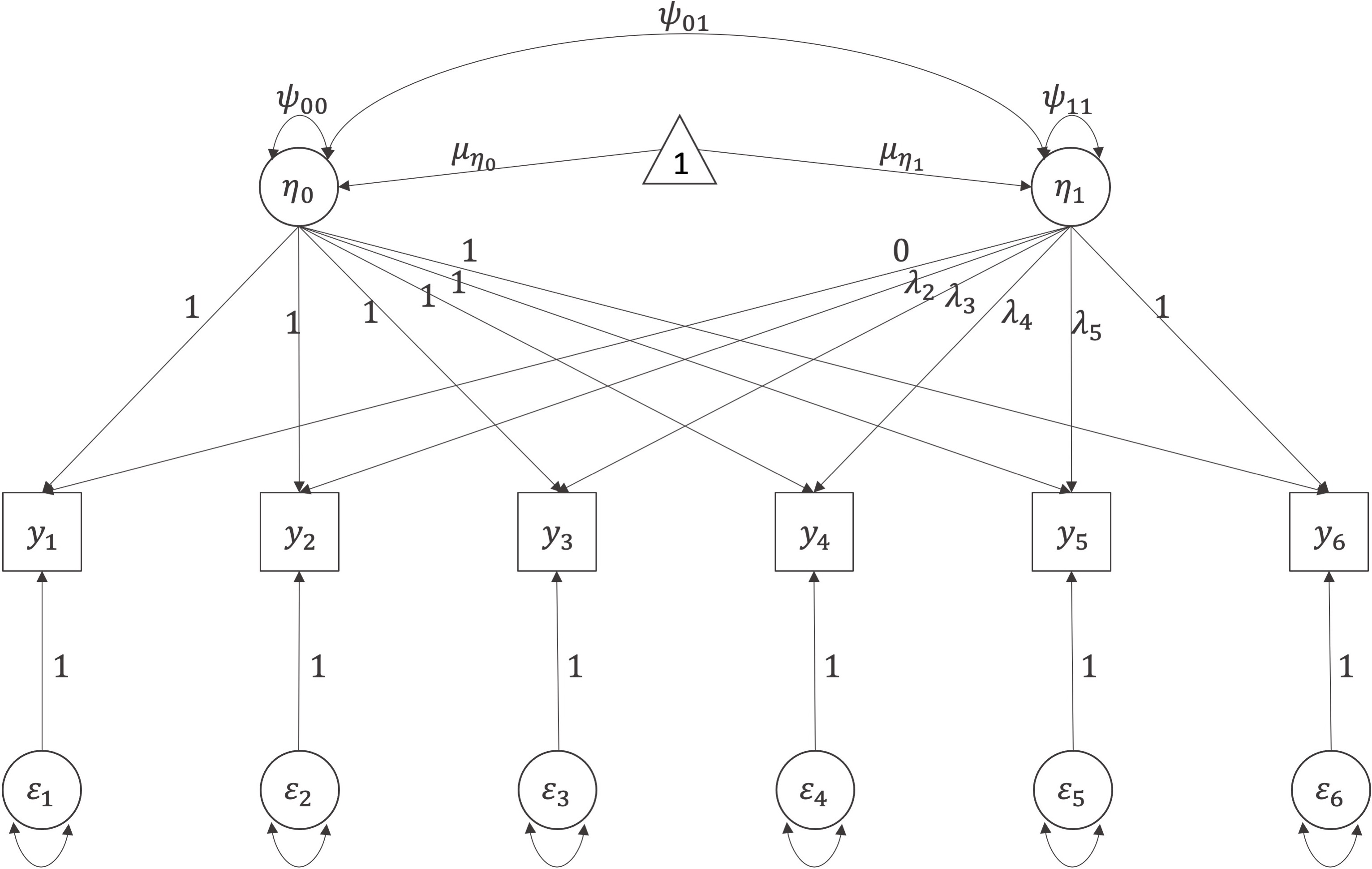}
 \caption{Specification 2}
 \label{fig:LBGM_old2}
\end{subfigure}
\caption{Path Diagram of Traditional Latent Basis Growth Models}
\label{fig:LBGM_old}
\end{figure}

\begin{figure}[!ht]
\centering
\begin{subfigure}{.5\textwidth}
 \centering
 \includegraphics[width=1.0\linewidth]{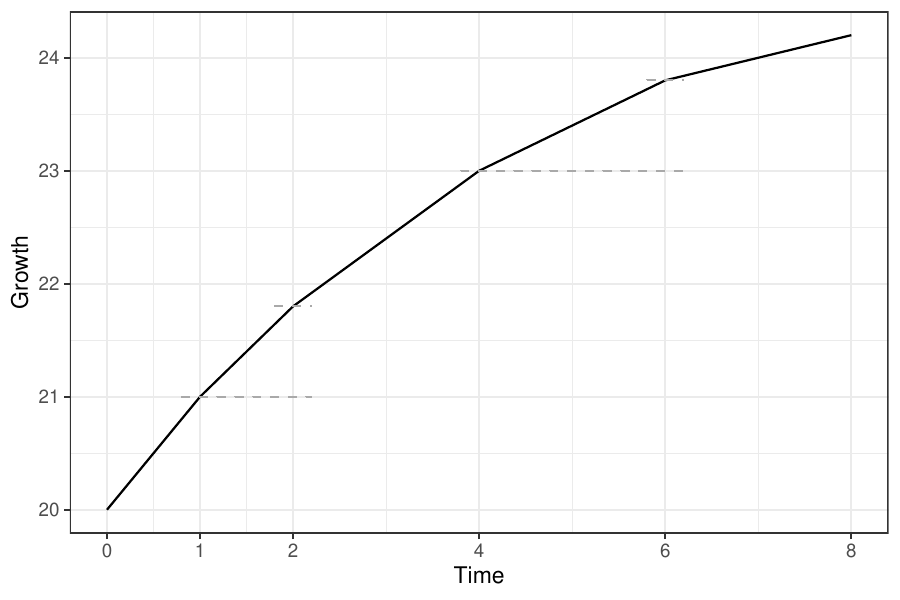}
 \caption{Piecewise Linear Growth Curve}
 \label{fig:LBGM_growth}
\end{subfigure}%
\begin{subfigure}{.5\textwidth}
 \centering
 \includegraphics[width=1.0\linewidth]{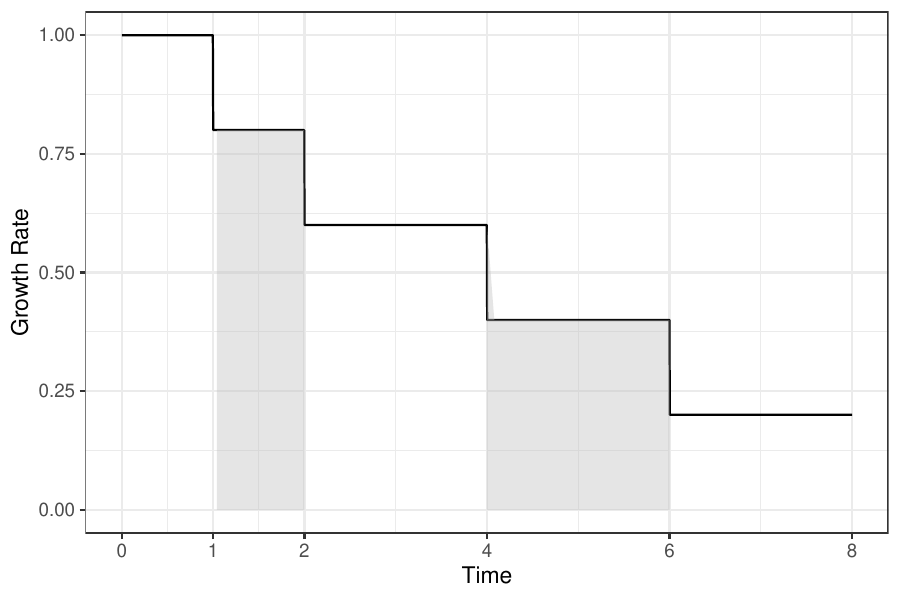}
 \caption{Piecewise Linear Growth Rate}
 \label{fig:LBGM_rate}
\end{subfigure}
\caption{Piecewise Linear Growth Curve and Growth Rate (Values of the Intercept and Slope of Each Time Interval: $\eta_{0}=20$; $\gamma_{1}=1.0$; $\gamma_{2}=0.8$; $\gamma_{3}=0.6$; $\gamma_{4}=0.4$; $\gamma_{5}=0.2$)}
\label{fig:LBGM}
\end{figure}

\begin{figure}[!ht]
\centering
\begin{subfigure}{.5\textwidth}
 \centering
 \includegraphics[width=1.0\linewidth]{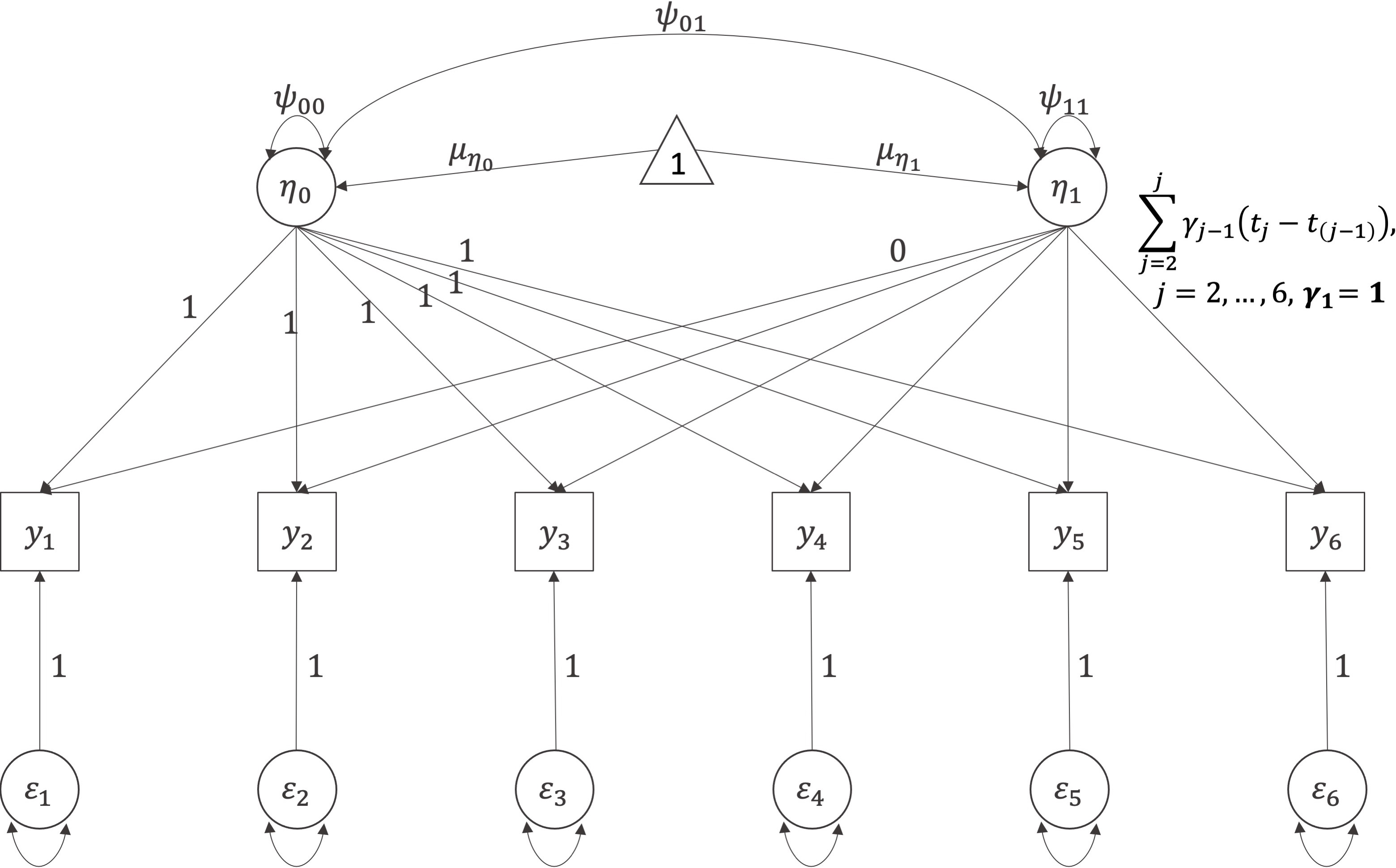}
 \caption{Specification 1}
 \label{fig:LBGM_new1}
\end{subfigure}%
\begin{subfigure}{.5\textwidth}
 \centering
 \includegraphics[width=1.0\linewidth]{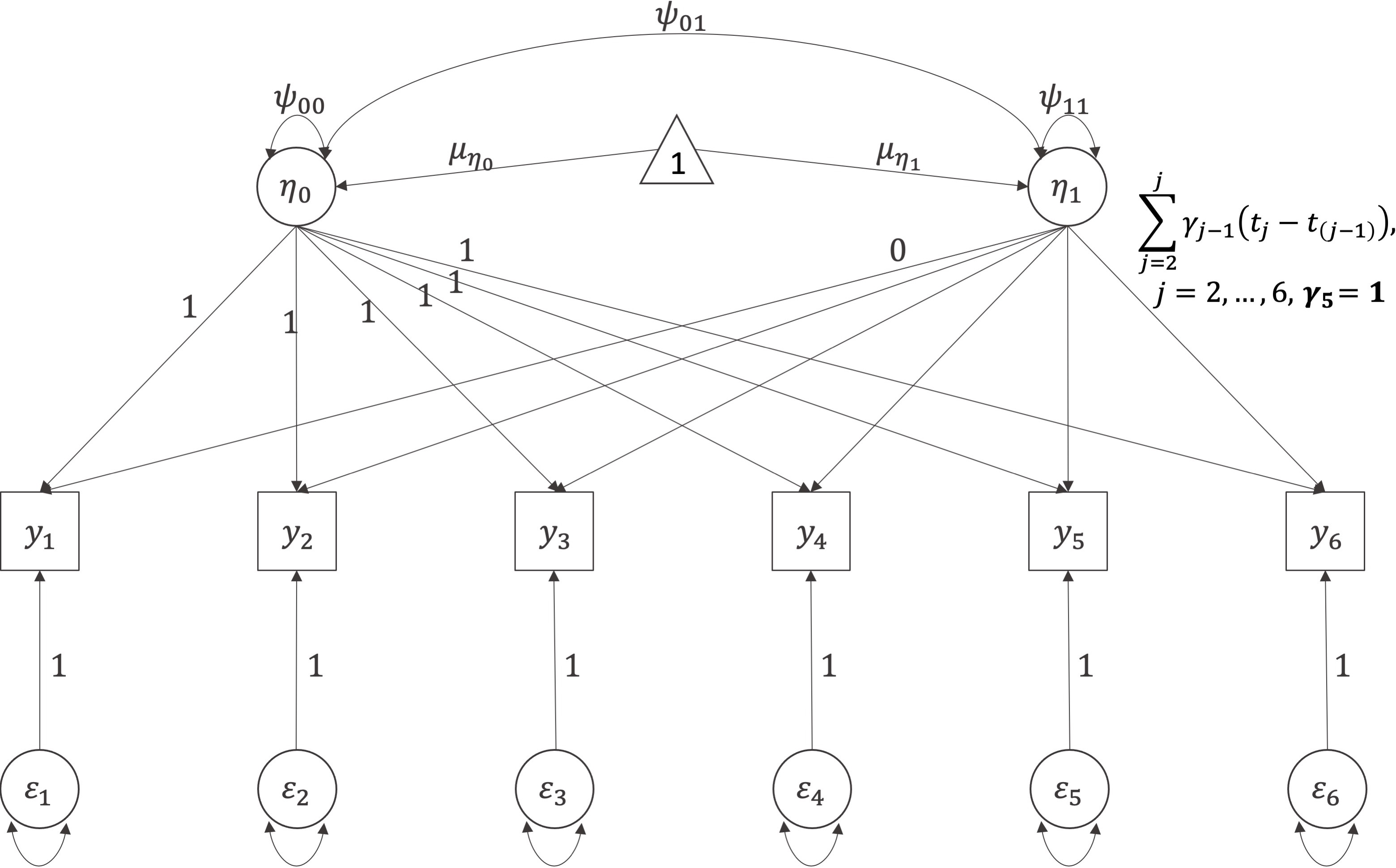}
 \caption{Specification 2}
 \label{fig:LBGM_new2}
\end{subfigure}
\caption{Path Diagram of Novel Latent Basis Growth Models}
\label{fig:LBGM_new}
\end{figure}

\begin{figure}[!ht]
\centering
\includegraphics[width=1.0\linewidth]{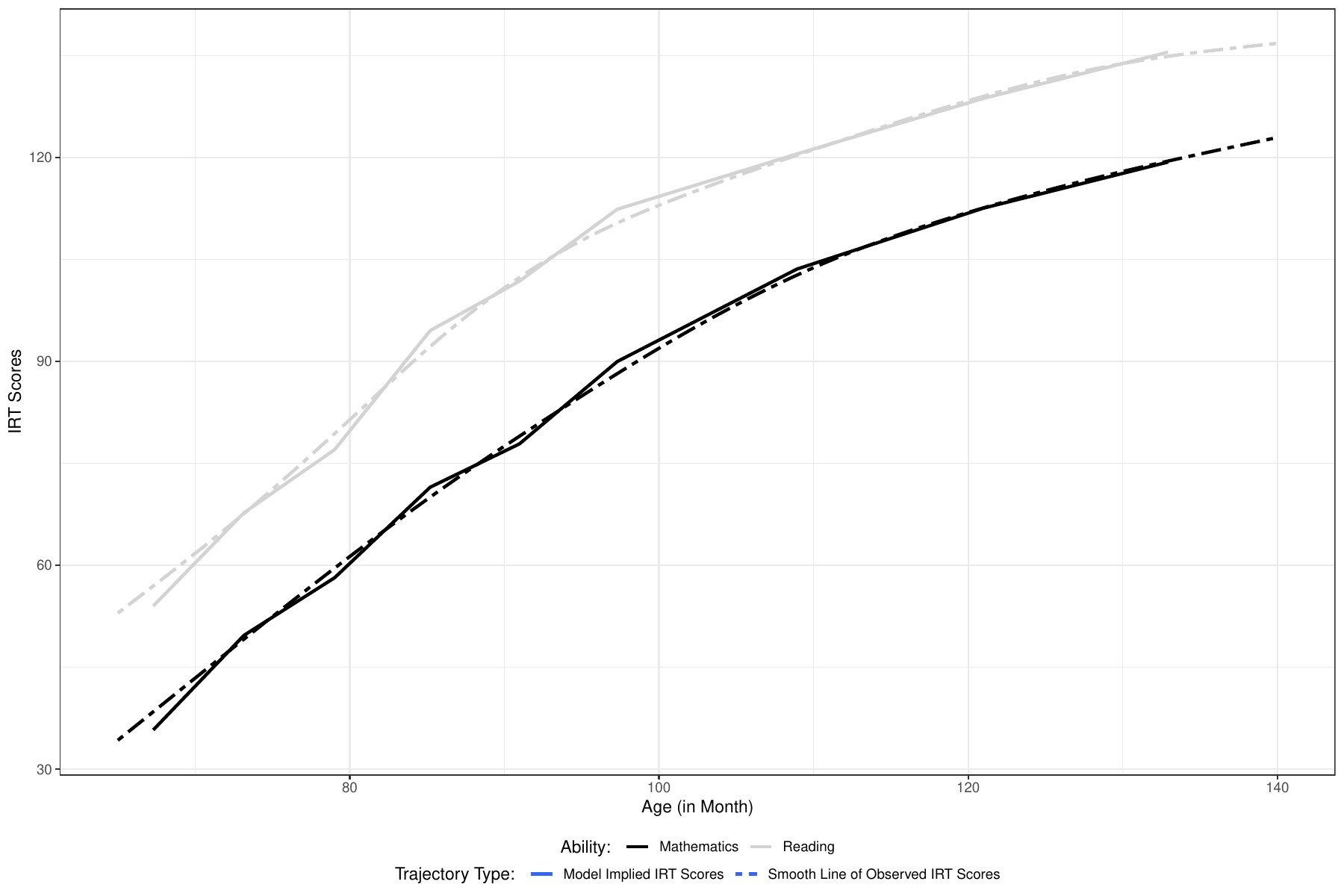}
\caption{Model Implied Trajectory and Smooth Line of Univariate Development}
\label{fig:Uni_traj}
\end{figure}

\begin{figure}[!ht]
\centering
\includegraphics[width=1.0\linewidth]{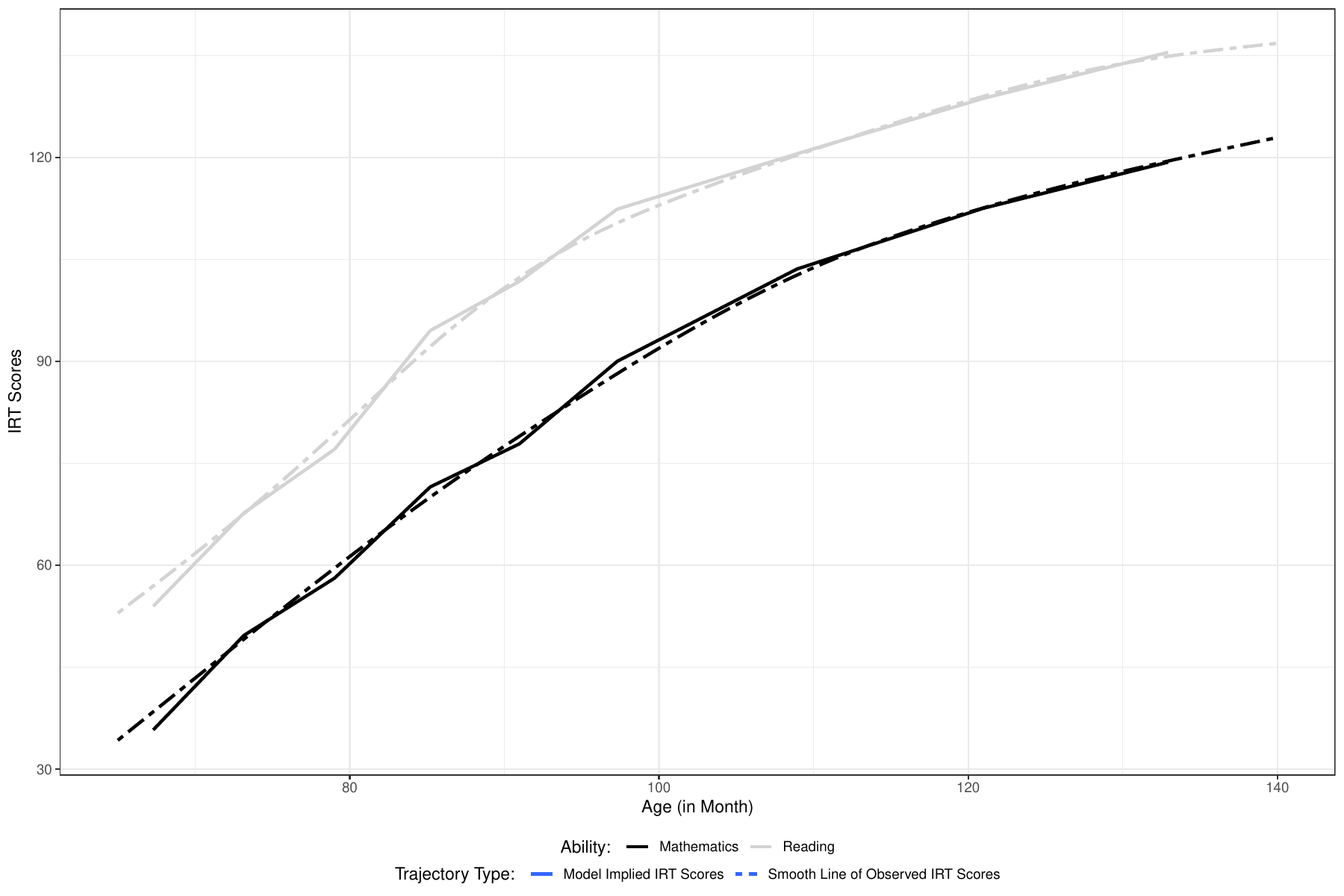}
\caption{Model Implied Trajectory and Smooth Line of Bivariate Development with The Same Time Structures}
\label{fig:Bi_traj}
\end{figure}

\begin{figure}[!ht]
\centering
\includegraphics[width=1.0\linewidth]{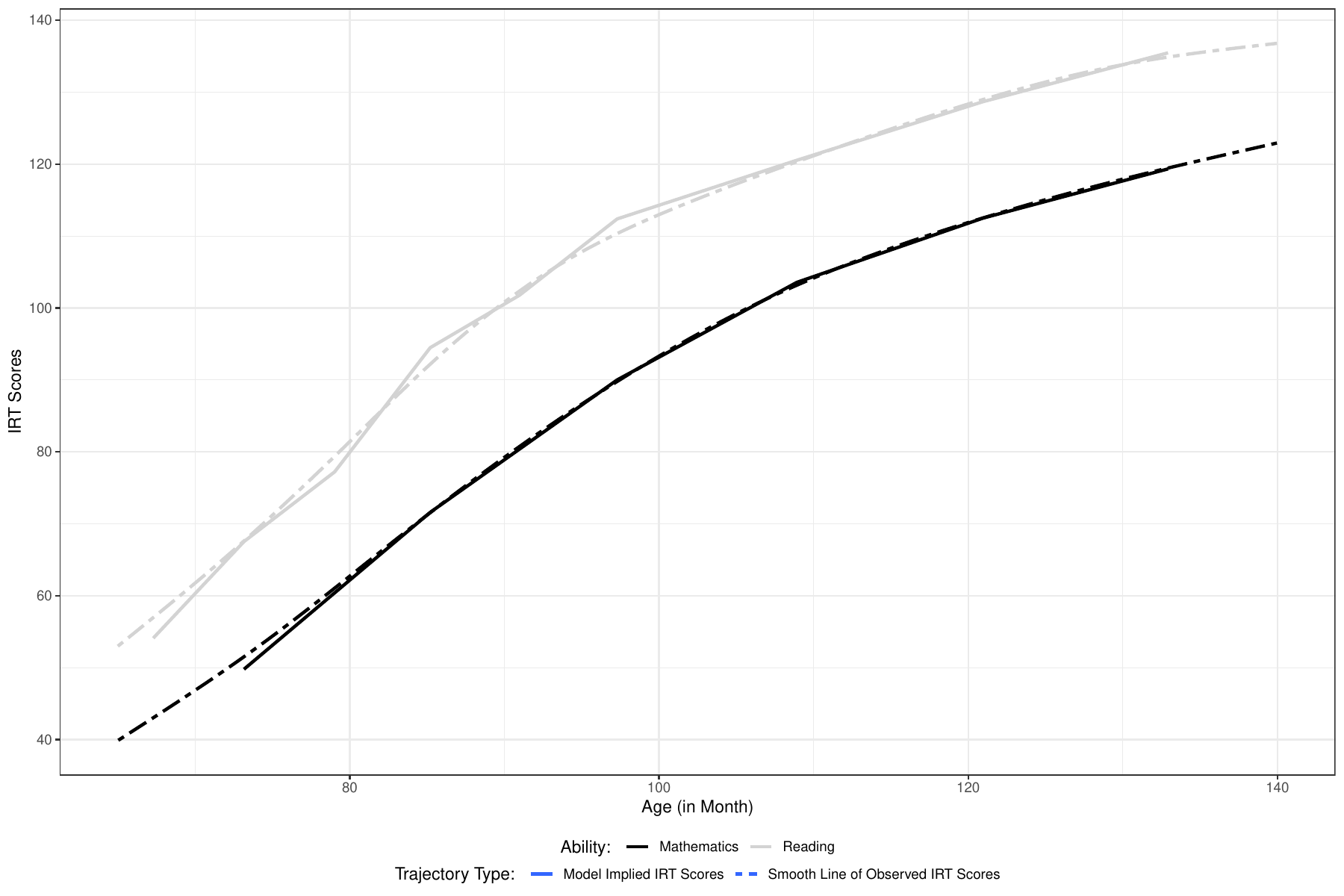}
\caption{Model Implied Trajectory and Smooth Line of Bivariate Development with Different Time Structures}
\label{fig:Bi_traj_diff}
\end{figure}

\newpage

\renewcommand\thetable{\arabic{table}}
\setcounter{table}{0}
\begin{table}[!ht]
\centering
\begin{threeparttable}
\caption{Performance Metrics: Definitions and Estimates}
\begin{tabular}{p{4cm}p{4.5cm}p{5.5cm}}
\hline
\hline
\textbf{Criteria} & \textbf{Definition} & \textbf{Estimate} \\
\hline
Relative Bias & $E_{\hat{\theta}}(\hat{\theta}-\theta)/\theta$ & $\sum_{s=1}^{S}(\hat{\theta}_{s}-\theta)/S\theta$ \\
Empirical SE & $\sqrt{Var(\hat{\theta})}$ & $\sqrt{\sum_{s=1}^{S}(\hat{\theta}_{s}-\bar{\theta})^{2}/(S-1)}$ \\
Relative RMSE & $\sqrt{E_{\hat{\theta}}(\hat{\theta}-\theta)^{2}}/\theta$ & $\sqrt{\sum_{s=1}^{S}(\hat{\theta}_{s}-\theta)^{2}/S}/\theta$ \\
Coverage Probability & $Pr(\hat{\theta}_{\text{lower}}\le\theta\le\hat{\theta}_{\text{upper}})$ & $\sum_{s=1}^{S}I(\hat{\theta}_{\text{lower},s}\le\theta\le\hat{\theta}_{\text{upper},s})/S$\\
\hline
\hline
\end{tabular}
\label{tbl:metric}
\begin{tablenotes}
\small
\item[a] {$\theta$: the population value of the parameter of interest}
\item[b] {$\hat{\theta}$: the estimate of $\theta$}
\item[c] {$S$: the number of replications and set as $1,000$ in our simulation study}
\item[d] {$s=1,\dots,S$: indexes the replications of the simulation}
\item[e] {$\hat{\theta}_{s}$: the estimate of $\theta$ from the $s^{th}$ replication}
\item[f] {$\bar{\theta}$: the mean of $\hat{\theta}_{s}$'s across replications}
\item[g] {$I()$: an indicator function}
\end{tablenotes}
\end{threeparttable}
\end{table}

\begin{table}[!ht]
\centering
\resizebox{1.15\textwidth}{!}{
\begin{threeparttable}
\setlength{\tabcolsep}{4pt}
\renewcommand{\arraystretch}{0.6}
\caption{Simulation Design for Parallel Latent Basis Growth Model in the Framework of Individual Measurement Occasions}
\begin{tabular}{p{7cm} p{14.5cm}}
\hline
\hline
\multicolumn{2}{c}{\textbf{Fixed Conditions}}\\
\hline
\textbf{Variables} & \textbf{Conditions} \\
\hline
Distribution of the Intercept & $\eta^{[y]}_{0i}\sim N(50, 5^{2})$; $\eta^{[z]}_{0i}\sim N(30, 5^{2})$ (i.e., $\mu^{[y]}_{\eta_{0}}=50$, $\psi^{[y]}_{00}=25$; $\mu^{[z]}_{\eta_{0}}=30$, $\psi^{[z]}_{00}=25$) \\
Distribution of the Shape Factor & $\eta^{[y]}_{1i}\sim N(4, 1^{2})$; $\eta^{[z]}_{1i}\sim N(5, 1^{2})$ (i.e., $\mu^{[y]}_{\eta_{1}}=4$, $\psi^{[y]}_{11}=1$; $\mu^{[z]}_{\eta_{1}}=5$, $\psi^{[z]}_{11}=1$) \\
Correlations of Within-construct GFs & $\rho^{[u]}=0.3$ ($u=y,z$) \\
Correlation between Residuals & $\rho_{\epsilon}=0.3$ \\
\hline
\hline
\multicolumn{2}{c}{\textbf{Manipulated Conditions (Full Factorial)}}\\
\hline
\textbf{Variables} & \textbf{Conditions} \\
\hline
Sample Size & $n=200, 500$\\
\hline
\multirow{3}{*}{Time ($t_{j}$)} & $6$ equally-spaced: $t_{j}=0, 1.00, 2.00, 3.00, 4.00, 5.00$\\
& $10$ equally-spaced: $t_{j}=0, 1.00, 2.00, 3.00, 4.00, 5.00, 6.00, 7.00, 8.00, 9.00$\\
& $10$ unequally-spaced: $t_{j}=0, 0.75, 1.50, 2.55, 3.00, 3.75, 4.50, 6.00, 7.50, 9.00$\\
\hline
Individual $t_{ij}$ & $t_{ij} \sim U(t_{j}-\Delta, t_{j}+\Delta)$ ($\Delta=0.25$) \\
\hline
\multirow{4}{*}{\makecell{Relative Growth Rate \\ in Each Time Interval\tnote{a}}} & $6$ waves: $r^{[u]}=1.0, 0.8, 0.6, 0.4, 0.2$ ($u=y,z$) \\
& $6$ waves: $r^{[u]}=0.2, 0.4, 0.6, 0.8, 1.0$ ($u=y,z$) \\
& $10$ waves: $r^{[u]}=1.0, 0.9, 0.8, 0.7, 0.6, 0.5, 0.4, 0.3, 0.2$ ($u=y,z$) \\
& $10$ waves: $r^{[u]}=0.2, 0.3, 0.4, 0.5, 0.6, 0.7, 0.8, 0.9, 1.0$ ($u=y,z$) \\
\hline
Correlation of Between-construct GFs & $\rho=0, \pm{0.3}$ \\ 
\hline
Residual Variance & $\theta^{[u]}_{\epsilon}=1, 2$ ($u=y,z$) \\
\hline
\hline
\end{tabular}
\label{tbl:simu_design}
\begin{tablenotes}
\small
\item[a] Growth rate is the relative growth rate, which is defined as the absolute growth rate over the value of shape factor. 
\end{tablenotes}
\end{threeparttable}}
\end{table}

\begin{table}[!ht]
\centering
\resizebox{1.15\textwidth}{!}{
\begin{threeparttable}
\setlength{\tabcolsep}{4pt}
\renewcommand{\arraystretch}{0.6}
\caption{Estimates of Parallel Latent Basis Growth Model for Reading and Mathematics Ability with the Same Time Structures}
\begin{tabular}{lrrrrrr}
\hline
\hline
& \multicolumn{2}{c}{\textbf{Reading IRT Scores}} & \multicolumn{2}{c}{\textbf{Math IRT Scores}} & \multicolumn{2}{c}{\textbf{Covariance}} \\
\hline
Mean & Estimate (SE) & P value & Estimate (SE) & P value & Estimate (SE) & P value \\
\hline
Initial Status\tnote{a} & $53.984$ ($0.724$) & $<0.0001^{\ast}\tnote{b}$ & $35.797$ ($0.659$) & $<0.0001^{\ast}$ & ---\tnote{c} & --- \\
Rate\tnote{d} of Interval $1$ & $2.341$ ($0.080$) & $<0.0001^{\ast}$ & $2.369$ ($0.071$) & $<0.0001^{\ast}$ & --- & --- \\
Rate of Interval $2$ & $1.596$ ($0.077$) & $<0.0001^{\ast}$ & $1.437$ ($0.068$) & $<0.0001^{\ast}$ & --- & --- \\
Rate of Interval $3$ & $2.823$ ($0.077$) & $<0.0001^{\ast}$ & $2.169$ ($0.066$) & $<0.0001^{\ast}$ & --- & --- \\
Rate of Interval $4$ & $1.256$ ($0.080$) & $<0.0001^{\ast}$ & $1.098$ ($0.070$) & $<0.0001^{\ast}$ & --- & --- \\
Rate of Interval $5$ & $1.679$ ($0.074$) & $<0.0001^{\ast}$ & $1.920$ ($0.066$) & $<0.0001^{\ast}$ & --- & --- \\
Rate of Interval $6$ & $0.701$ ($0.041$) & $<0.0001^{\ast}$ & $1.167$ ($0.036$) & $<0.0001^{\ast}$ & --- & --- \\
Rate of Interval $7$ & $0.676$ ($0.039$) & $<0.0001^{\ast}$ & $0.742$ ($0.034$) & $<0.0001^{\ast}$ & --- & --- \\
Rate of Interval $8$ & $0.566$ ($0.040$) & $<0.0001^{\ast}$ & $0.568$ ($0.034$) & $<0.0001^{\ast}$ & --- & --- \\
\hline
\hline
Variance & Estimate (SE) & P value & Estimate (SE) & P value & Estimate (SE) & P value \\
\hline
Initial Status & $164.926$ ($13.085$) & $<0.0001^{\ast}$ & $139.878$ ($10.876$) & $<0.0001^{\ast}$ & $126.794$ ($10.650$) & $<0.0001^{\ast}$ \\
Rate of Interval $1$ & $0.111$ ($0.013$) & $<0.0001^{\ast}$ & $0.106$ ($0.011$) & $<0.0001^{\ast}$ & $0.064$ ($0.009$) & $<0.0001^{\ast}$ \\
Rate of Interval $2$ & $0.051$ ($0.007$) & $<0.0001^{\ast}$ & $0.039$ ($0.005$) & $<0.0001^{\ast}$ & $0.026$ ($0.004$) & $<0.0001^{\ast}$ \\
Rate of Interval $3$ & $0.161$ ($0.018$) & $<0.0001^{\ast}$ & $0.089$ ($0.010$) & $<0.0001^{\ast}$ & $0.070$ ($0.009$) & $<0.0001^{\ast}$ \\
Rate of Interval $4$ & $0.032$ ($0.005$) & $<0.0001^{\ast}$ & $0.023$ ($0.003$) & $<0.0001^{\ast}$ & $0.016$ ($0.002$) & $<0.0001^{\ast}$ \\
Rate of Interval $5$ & $0.057$ ($0.007$) & $<0.0001^{\ast}$ & $0.070$ ($0.008$) & $<0.0001^{\ast}$ & $0.037$ ($0.005$) & $<0.0001^{\ast}$ \\
Rate of Interval $6$ & $0.010$ ($0.001$) & $<0.0001^{\ast}$ & $0.026$ ($0.003$) & $<0.0001^{\ast}$ & $0.009$ ($0.001$) & $<0.0001^{\ast}$ \\
Rate of Interval $7$ & $0.009$ ($0.001$) & $<0.0001^{\ast}$ & $0.010$ ($0.001$) & $<0.0001^{\ast}$ & $0.006$ ($0.001$) & $<0.0001^{\ast}$ \\
Rate of Interval $8$ & $0.006$ ($0.001$) & $<0.0001^{\ast}$ & $0.006$ ($0.001$) & $<0.0001^{\ast}$ & $0.004$ ($0.001$) & $0.0001^{\ast}$ \\
\hline
\hline
\end{tabular}
\label{tbl:RM_est}
\begin{tablenotes}
\small
\item[a] The initial Status was defined as $60$ months old in this case.\\
\item[b] $^{\ast}$ indicates statistical significance at $0.05$ level.\\
\item[c] --- indicates that the metric was not available in the model. \\
\item[d] The mean, variance, and covariance of rate in each interval were the corresponding value of absolute growth rate, which can be obtained by \textit{R} function \textit{mxAlgebra()} from estimated shape factor and relative growth rate.
\end{tablenotes}
\end{threeparttable}}
\end{table}

\begin{table}[!ht]
\centering
\resizebox{1.15\textwidth}{!}{
\begin{threeparttable}
\setlength{\tabcolsep}{4pt}
\renewcommand{\arraystretch}{0.6}
\caption{Estimates of Parallel Latent Basis Growth Model for Reading and Mathematics Ability with Different Time Structures\tnote{a}}
\begin{tabular}{lrrrrrr}
\hline
\hline
& \multicolumn{2}{c}{\textbf{Reading IRT Scores}} & \multicolumn{2}{c}{\textbf{Math IRT Scores}} & \multicolumn{2}{c}{\textbf{Covariance}} \\
\hline
Mean & Estimate (SE) & P value & Estimate (SE) & P value & Estimate (SE) & P value \\
\hline
Initial Status\tnote{b} & $54.100$ ($0.724$) & $<0.0001^{\ast}\tnote{c}$ & $49.782$ ($0.694$) & $<0.0001^{\ast}$ & ---\tnote{d} & --- \\
Rate\tnote{e} of Interval\tnote{f} $1$ & $2.288$ ($0.080$) & $<0.0001^{\ast}$ & --- & --- & --- & --- \\
Rate of Interval $2$ & $1.653$ ($0.077$) & $<0.0001^{\ast}$ & $1.811$ ($0.037$) & $<0.0001^{\ast}$ & --- & --- \\
Rate of Interval $3$ & $2.793$ ($0.077$) & $<0.0001^{\ast}$ & $1.811$ ($0.037$) & $<0.0001^{\ast}$ & --- & --- \\
Rate of Interval $4$ & $1.263$ ($0.080$) & $<0.0001^{\ast}$ & $1.523$ ($0.036$) & $<0.0001^{\ast}$ & --- & --- \\
Rate of Interval $5$ & $1.679$ ($0.074$) & $<0.0001^{\ast}$ & $1.523$ ($0.036$) & $<0.0001^{\ast}$ & --- & --- \\
Rate of Interval $6$ & $0.702$ ($0.041$) & $<0.0001^{\ast}$ & $1.167$ ($0.037$) & $<0.0001^{\ast}$ & --- & --- \\
Rate of Interval $7$ & $0.676$ ($0.039$) & $<0.0001^{\ast}$ & $0.742$ ($0.035$) & $<0.0001^{\ast}$ & --- & --- \\
Rate of Interval $8$ & $0.567$ ($0.039$) & $<0.0001^{\ast}$ & $0.572$ ($0.035$) & $<0.0001^{\ast}$ & --- & --- \\
\hline
\hline
Variance & Estimate (SE) & P value & Estimate (SE) & P value & Estimate (SE) & P value \\
\hline
Initial Status & $164.755$ ($13.095$) & $<0.0001^{\ast}$ & $156.78$ ($12.804$) & $<0.0001^{\ast}$ & $130.957$ ($11.353$) & $<0.0001^{\ast}$ \\
Rate of Interval $1$ & $0.105$ ($0.012$) & $<0.0001^{\ast}$ & --- & --- & --- & --- \\
Rate of Interval $2$ & $0.055$ ($0.007$) & $<0.0001^{\ast}$ & $0.060$ ($0.007$) & $<0.0001^{\ast}$ & $0.032$ ($0.005$) & $<0.0001^{\ast}$ \\
Rate of Interval $3$ & $0.157$ ($0.018$) & $<0.0001^{\ast}$ & $0.060$ ($0.007$) & $<0.0001^{\ast}$ & $0.055$ ($0.008$) & $<0.0001^{\ast}$ \\
Rate of Interval $4$ & $0.032$ ($0.005$) & $<0.0001^{\ast}$ & $0.042$ ($0.005$) & $<0.0001^{\ast}$ & $0.021$ ($0.003$) & $<0.0001^{\ast}$ \\
Rate of Interval $5$ & $0.057$ ($0.007$) & $<0.0001^{\ast}$ & $0.042$ ($0.005$) & $<0.0001^{\ast}$ & $0.028$ ($0.004$) & $<0.0001^{\ast}$ \\
Rate of Interval $6$ & $0.010$ ($0.001$) & $<0.0001^{\ast}$ & $0.025$ ($0.003$) & $<0.0001^{\ast}$ & $0.009$ ($0.001$) & $<0.0001^{\ast}$ \\
Rate of Interval $7$ & $0.009$ ($0.001$) & $<0.0001^{\ast}$ & $0.010$ ($0.001$) & $<0.0001^{\ast}$ & $0.005$ ($0.001$) & $<0.0001^{\ast}$ \\
Rate of Interval $8$ & $0.006$ ($0.001$) & $<0.0001^{\ast}$ & $0.006$ ($0.001$) & $<0.0001^{\ast}$ & $0.004$ ($0.001$) & $0.0001^{\ast}$ \\
\hline
\hline
\end{tabular}
\label{tbl:RM_est_diff}
\begin{tablenotes}
\small
\item[a] In this joint model, we included the measurements of reading ability at all nine waves, but we only included the measures of mathematics ability at Wave $2$, $4$, $6$, $7$, $8$, and $9$.
\item[b] The initial Status of reading ability was defined as the measurement at $60$ months old, while that of mathematics ability was the measurement half a year later. 
\item[c] $^{\ast}$ indicates statistical significance at $0.05$ level.
\item[d] --- indicates that the metric was not available in the model. 
\item[e] The mean, variance, and covariance of rate in each interval were the corresponding value of absolute growth rate, which can be obtained by \textit{R} function \textit{mxAlgebra()} from estimated shape factor and relative growth rate.
\item[f] Each `interval' was defined as the interval between any two consecutive measurement occasions of reading ability. The estimates of mathematics ability during the first interval are not applicable because the first measure of mathematics ability was Wave $2$. The estimated means and variances of mathematics ability in Interval 2 (Interval 4) and Interval 3 (Interval 5) were the same because we took out its measurement at Wave $3$ (Wave $5$).
\end{tablenotes}
\end{threeparttable}}
\end{table}

\end{document}